Short Paper

# Investigating the Influence of Computer Anxiety on the Academic Performance of Junior Secondary School Students in Computer Studies in Nigeria

Chinelo Blessing Oribhabor
Department of Guidance and Counseling, Faculty of Arts and Education, University of Africa

**Abstract**

*Purpose* – This study examined the influence of computer anxiety on the academic performance of junior secondary school students (JSS) in Computer Studies in Nigeria. The study specifically examined whether there is any correlation between the computer anxiety of students and their performance in Computer Studies.

*Method* – The sample consisted of one thousand, two hundred JSS 3 students from sixty selected secondary schools in the 12 selected states from each of the six geopolitical zones in Nigeria. The research instrument that was used in the study was computer anxiety scale which was validated by the Guidance and Counseling lecturers and Educational Measurement and Evaluation experts. Cronbach's alpha technique was used to determine the reliability of the computer anxiety scale and 0.87 coefficient was achieved. The students' second term results of the 2018/ 2019 session were used to measure their academic performance in Computer Studies. Independent t-test and Pearson Product Moment Correlation were used to analyze the data collected.

*Results* – The result of the study showed that most of the students used in the study were mildly anxious when dealing with computer; there is no significant difference in the computer anxiety of male and female junior secondary school students in Nigeria; and there is a significant negative relationship between computer anxiety and Computer Studies performance of junior secondary school students' in Nigeria.

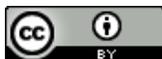

*Conclusion* – The study concludes that students' anxious feelings of operating computer do negatively affect their performances in the computer studies.

*Recommendation* – Based on the findings, the researcher recommends amongst others that Computer Studies' teachers should indulge the students in constant use of computer for assignments and web-browsing so as to enhance their confidence and reduce anxiety in Computer Studies classroom.

*Practical Implication* – The results of the study reflect the need for the Nigerian Government to provide enough computer sets to schools so that each students will be entitled to a computer during their training.

## INTRODUCTION

Development in the area of technology is a yardstick used in measuring a nation's development. Technological growth of a nation leads to its social and economic development. In the world today, science and technology has become a dominant power development indicator. China, America, Germany, Japan and Russia are typical examples of nations which are now referred to as developed, as a result of their development in the area of science and technology. At the heart of science and technology is information communication technology (ICT). In the society nowadays, people can assess information through the use of computer technology. The computer is a technological innovation under the control of stored programme that can perform some of the intellectual roles of man even beyond human capability. It is a power-driven machine equipped with keyboards, electronic circuits, storage compartments, and recording devices for the high speed performance of mathematical operations.

Adigun et al. (2015) defined computer science as the study of the theory, design, use and analysis of computer devices. This entails knowing the computer itself, its operation, what it can do, how it can do it and why it's doing it, these form the basis of computer studies curriculum in secondary school. Computers are used not only to manage information and to enhance productivity, but also used in education. As the world is going computerized, people need to be literate in operating computers so as to function effectively and efficiently at work place and school. Computers have been accepted unreservedly as an integral to the education system. The students of today are to apply the skills they learnt during academic programs in their respective fields in future. They would be required to use computer, the internet and its various services, work on the word processing, databases, spreadsheets, software, power-point slides, and so on. Ogwo, Maidoh and Onwe (2015) stated that the advantages of Computer Education in schools are for information storage, audio-visual aids in teaching, enhancement of communication among students, teachers and parents, easy, effective and better presentation of information.

In Nigeria, the Federal Government saw the need of computer in schools and introduced Computer Education into the nation's secondary school system in August 1988 through the policy enactment of the National Computer Policy in which the objectives are to ensure literacy in computer education at the secondary school level; and help meet with the demand of our time and technological development. To achieve the objectives of introducing Computer Education in secondary school curriculum, the Federal Ministry of Education in Nigeria had spent a great amount of money on Computer studies by providing computer sets, accessories and also launched an Information Communication Technology (ICT)-driven project which is known as School Net (Adomi & Kpangban, 2010). The introduction of Computer Studies in the secondary schools' curriculum in Nigeria has paved way for students' exposure to information Communication Technology (ICT). As there is a growing demand on the use of computers in secondary schools, a great number of students are happy and eager to learn how to operate computers while some are afraid to operating computers. Hence, despite having basic skills and knowledge in computer, some students feel anxious while using computer to accomplish their tasks, as fluent interaction with computer interfaces requires both the basic understanding of computer concepts as well as the basic familiarity with hardware and software.

The anxiousness about operating computer gives serious problem to some students and this can affect their performance in the Computer Studies in school. Some researchers (e.g., Iniobong, 2018; Adebowale, Adediwura & Balla, 2009) revealed that a large proportion of students fail Computer Science in their internal and external examinations, and as a result, they hardly attain the acceptable level of good performance in their results. It has been claimed that the stumbling blocks has always been their inability to master and pass Computer Science. Some researchers (Ayodele & Adebiyi, 2013; Oladimeji et al., 2018) also have it that the causes of poor performance in Computer Science in the Junior Certificate Examination are lack of trained teachers in Computer science, lack of proper teaching method in Computer Science, non-use of instructional materials in teaching Computer Science but much research has not been done on the anxiety as one of the causes of poor performance of students in Computer science examination, especially in Nigeria. Shah, Hassan and Embi (2012) opined that students' anxiety can occur when there is a complication in the use of computers or when the complication application is too advanced to cope with or when the system is down or malfunctioning.

The present study attempts to give an insight to the computer anxiety levels of students and their performances in Computer Studies. It examines the relationship between computer anxiety and academic performance in Computer Studies and a background characteristic such as gender. It is noteworthy that there is no previous study in Nigeria to discuss computer anxiety and students Performance in Computer Studies at the same time. This study is significant in that it will add to the literature regarding computer anxiety and performance in Computer Studies of Nigerian junior secondary school students.

**LITERATURE REVIEW**

Anxiety refers to a complex combination of negative emotional responses that include worry, fear, apprehension and agitation (Saadé & Kira, 2009). Olatoye (2009) stated that anxiety is an intense dread, apprehension or nagging worry while Mathew (2012) explained anxiety is a natural and unavoidable reaction to a perception of danger or risk. Simsek (2011) added that computer anxiety is considered as an effective response while Laosethakul and Leingpibul (2010) explained that the feelings of anxiety may be mediated by beliefs about lack of ability to use a computer knotted to a lack of mathematical and mechanical skills. Kannan, Muthumanickam and Chandrasekaran (2016) refer computer anxiety as when a student is afraid, uneasy to use computer. An individual is said to be computer anxious when the person's state of emotion during interaction with computer decreases the advantages which one can get as a result of using computer. According to Torkzadeh and Angulo (1992) cited in Simsek (2011), there are three major dimensions of computer anxiety as psychological, operational, and sociological. To be more concrete, psychological dimension includes attitudes toward computers, self-efficacy, personality types, avoidance, and self-perceptions. Operational dimension usually results from computer courses, teachers, nature of computers, the extent of experiences with the computer, and owning a personal computer. Sociological dimension is related to factors of age, gender, nationality, socio-economic status, and the field of study.

Computer anxiety is a substantial barrier influencing the use of computers ultimately influencing the academic activities of students. A person with computer anxiety may experience fear of the unknown, feeling of frustration, possible embarrassment, failure and disappointment which may result to avoidance towards computer usage (Olatoye, 2009). Simsek (2011) also explained that computer anxiety is considered as an effective response and that the feelings of anxiety may be mediated by beliefs about lack of ability to use a computer knotted to a lack of mathematical and mechanical skills. Abele and Spurk (2009) opined that computer anxiety has influenced an individual's choice of learning about computers and achieving a realistic level of competency in computer usage. Shermis and Lombart (1998), in their study, suggested that much of what was considered computer anxiety might in fact be a manifestation of test anxiety. Al-Khaldi and Al-Jabri (1998) found that the strongest predictor of computer utilization was computer liking followed by confidence. Researchers such as Tekinarslan (2008), Chang (2005) and Roussos (2004), found that a variable that appeared to have a strong influence on computer anxiety was computer experience. Though, Anderson (1996) cited in Korobili, Togia & Malliari (2010) concluded that perceived knowledge rather than experience was a predictor of computer anxiety. Glaister (2007) found that people who reported to have medium to high levels of computer anxiety performed less well than those with low level of computer anxiety in an examination involving the use of computers. The higher the anxiety in operating computers, the higher is the tendency of committing academic procrastination (Rahardjo & Juneman, 2013). However, Singh, Chandwani, Singh and Kumar, (2019) found that computer anxiety was negatively correlated with student's grade point, experience in using computer and time spent on computer and internet.

Gender is one of factors also mentioned in literature to have considerable effects on students' academic performances especially in science subjects. Gender is one of such factors also mentioned in literature to have considerable effects on students' academic performances. Gender is the range of physical, biological, mental and behavioural characteristics pertaining to and differentiating between the feminine and masculine (female and male) population. The importance of examining performance in relation to gender is based primarily on the socio-cultural differences between males and females. Some vocations and professions have been regarded as men's (engineering, arts and crafts, agriculture etc.) while others as women's (catering, typing, nursing etc.). As a result of this way of thinking, the larger society has tended to see girls as a weaker sex. Consequently, an average Nigerian girl goes to school with these fixed stereotypes. Gender differences in achievement have been examined for some time resulting in a substantial body of literature (Adeyemi & Ajibade, 2011; Dania, 2014; Agbaje & Alake, 2014; Atovigba, Vershima, O'Kwu & Ijenkeli, 2012). Some of these researchers pointed out that there is no significant gender difference in students' academic achievement while others found significant difference with either the males or the females performing better. Moreover, some studies found that females are more Computer anxious than males (Chou, 2003; Durndell & Haag, 2002), while other researchers found that there is no significant differences between males and females (Sam, Othman & Nordin, 2005; Roussos, 2004). Based on the literature review, it can be concluded that research studies regarding gender differences in affective variable such as anxiety are inconclusive and need more attention. In order to fill in these gaps, this research paper attempts to explore the influence of computer anxiety on junior secondary school academic performance in Computer Studies.

**PURPOSE OF THE STUDY**

The main purpose of this study was to investigate the influence of computer anxiety on the academic performance of junior secondary school students in Computer Studies in Nigeria.

**RESEARCH QUESTIONS**

The study provides answers to the following questions:
i. What is the Computer Studies performance of junior secondary school students in Nigeria?
ii. What is the level of Computer anxiety in junior secondary school students in Nigeria?
iii. Is there difference in the Computer anxiety of male and female junior secondary school students in Nigeria?
iv. Is there relationship between computer anxiety and Computer Studies performance of the junior secondary school students' in Nigeria?

**RESEARCH HYPOTHESES**

The following hypotheses were tested at 0.05 significance level:
i. There is no significant difference in the computer anxiety of male and female junior secondary school students in Nigeria.
ii. There is no significant relationship between computer anxiety and Computer Studies performance of junior secondary school students' in Nigeria.

**METHODOLOGY**

The study is a research conducted through a descriptive survey. The population for the study comprised all the students in the junior secondary school three (JSS3) in Nigeria. Multi-stage sampling technique was used in the study. First, stratified sampling technique was used to divide the states in Nigeria into geographical zones. There are six geographical zones in Nigeria: North Central, North-East, North-West, South-East, South-South and South-West. Second, simple random sampling technique was used to select two states each from the six geopolitical zones, which results to 12 states used in the study (Niger and Plateau states from North Central; Adamawa and Taraba States from North-East; Kaduna and Katsina States from North-West; Abia and Enugu States from South-East; Edo and Delta States from South-South; Lagos and Ekiti States from South-West). Third, purposive sampling techniques were used to select five schools each from the 12 randomly selected states, totaling 60 schools used in the study.

The schools were purposively selected because they are the schools that offer Computer Studies in their various schools. Finally, simple random sampling technique was used to select 20 JSS 3 students from each of the 60 selected schools, totaling 1,200 students used as sample in the study. The questionnaire used to gather data from the students was an adopted computer anxiety scale by Cohen & Waugh (1989). The Computer Anxiety Scale consisted of 16 items and were scored from 1 to 5 with "1" indicating a response of "strongly disagree" to "5" indicating a response of "strongly agree". Sample of the questions in the questionnaire are: 'I feel anxious whenever I am using computers'; 'I am confident in my ability to use computers'; 'I feel relaxed when I am working on a computer'; and 'I am frightened by computers'. The Computer Anxiety

Scale was computed to get a total of computer anxiety scores of students and was grouped into five levels of computer anxiety. Higher scores indicated as high level of computer anxiety while low score indicated low level of computer anxiety. The minimum and maximum scores range from 16 to 80. These ranged from (a) very relaxed, 16 to 28; (b) generally relaxed, 29 to 41; (c) mildly anxious, 42 to 54; (d) anxious, 55 to 67; (e) very anxious, 68 to 80. Cronbach alpha technique was used to determine the reliability of the computer anxiety scale and 0.87 coefficient was achieved. Validity of the computer anxiety scale was ensured by the Guidance and Counseling lecturers and Educational Measurement and Evaluation expert judgments. To measure Computer Studies performance of the students in the study, the researcher collected the students' second term results (i.e., grades) of the 2018/ 2019 academic year. The reason why the researcher used the students' second term results (grades) was because the JSS 3 students' third term examination is written as external examination (National Examination- Basic Education Certificate Examination) in which the researcher may not be able to get access to the results. The Computer Studies performance scores of the students were standardized using T-score so as to make the scores of the students to be uniform. Descriptive statistical analysis was done using frequencies, percentages, mean and standard deviation while the inferential statistical analysis involved using independent t-test and Pearson Product Moment correlation. Hypotheses were tested at the 0.05 level of significance.

**RESULTS**

**Research Question One:** What is the Computer Studies performance of junior secondary school students in Nigeria?

Table 1 shows the Computer Studies performance of junior secondary school students in the selected schools in Nigeria. Table 1 shows that 582 students (48.5%) used in the study have C grade (lower credit), followed by 301 students (25.1%) who have B grade (upper credit), followed by 274 students (22.83%) who have P grade (pass), followed by 41 students (3.4%) who failed and the least is A grade which was made by 2 students (0.17%). The result in Table 1 shows that a considerable number of students have lower credit in Computer Studies examination in their schools.

Table 1. Description of JSS 3 Students' Performance in the 2018/ 2019 Second Term Computer Studies Examinations in the Selected Schools

| Grades | Definition | Frequency | Percentage |
|---|---|---|---|
| A | Distinction | 2 | 0.17% |
| B | Upper Credit | 301 | 25.1% |
| C | Lower Credit | 582 | 48.5% |
| P | Pass | 274 | 22.83% |
| F | Fail | 41 | 3.4% |
| Total | | 1,200 | 100% |

**Research Question Two:** What is the level of Computer anxiety in junior schools in Nigeria?

Table 2 shows the level of computer anxiety of the JSS 3 students in the selected schools in Nigeria. There are only 1.42 % (n=17) who shows very relaxed among the students. The table 2 also shows that some students have generally relaxed by 13.25% (n= 159), while the highest percentage was mildly anxious with 60.33% (n=724). Moreover, 20.67% (n= 248) are anxious and 4.33% (n= 52) are very anxious. The mean computer anxiety scores is 2.94 and standard deviation scores is 0.27. Based on the findings, most of the students used in the study were mildly anxious when dealing with computer.

**Hypothesis One:** There is no significant difference in the computer anxiety of male and female junior secondary school students in Nigeria.

Table 3 shows the male respondents' mean= 22.34; standard deviation= 3.29 and the female respondents' mean= 21.88; standard deviation= 3.27; t= 0.204 and p-value= 0.643. Testing the hypothesis at the alpha level of 0.05, the p-value is greater than the alpha value, this shows that there is no significant difference; hence the null hypothesis is retained. Therefore, there is no significant difference in the computer anxiety of male and female junior secondary school students in Nigeria.

Table 2. Description of the Students' Computer Anxiety Levels

| Anxiety Level | Frequency | Percentage |
|---|---|---|
| Very Relaxed | 17 | 1.42% |
| Generally Relaxed | 159 | 13.25% |
| Mildly Anxious | 724 | 60.33% |
| Anxious | 248 | 20.67% |
| Very Anxious | 52 | 4.33% |
| **Total** | **1,200** | **100%** |

Table 3. Independent t-test of the difference in the Computer Anxiety of male and female Students in the Selected Schools Nigeria

| Variable | N | Mean | Standard Deviation | df | t-value | p-value |
|---|---|---|---|---|---|---|
| Male | 476 | 22.34 | 3.29 | 1198 | 0.204 | 0.643 |
| Female | 724 | 21.88 | 3.27 | | | |

**Hypothesis Two:** There is no significant relationship between computer anxiety and Computer Studies performance of junior secondary school students' in Nigeria.

Table 4 shows the relationship between computer anxiety and Computer Studies performance of the selected junior secondary school students' in Nigeria. The Table 4 shows the correlation coefficient or r = -0.432, this means that there is a negative correlation. The Table 4 also shows p-value = 0.001, testing the hypothesis at 0.05 alpha level, the p-value is less than the alpha value, which shows a significant relationship. Therefore, the null hypothesis is rejected. This means that there is a significant negative relationship between computer anxiety and Computer Studies performance of junior secondary school students' in Nigeria.

Table 4. Pearson Product Moment Correlation of Computer Anxiety and Computer Studies Performance of junior Secondary School Students' in Nigeria.

| Variable | N | r | p-value |
|---|---|---|---|
| Computer Anxiety Score | 1200 | -0.432 | 0.001 |
| Students' Computer Studies Performance | | | |

**DISCUSSION OF FINDINGS**

The finding of this study revealed that 724 (60.33%) students out of 1,200 students used in the study were mildly anxious when dealing with computer. This means that the students' anxiousness is moderate and that may be the reason why they performed moderately in their examination in which 582 students (60.33%) have lower credit results. This result is supported by Simsek (2011), who examined the relationship between computer anxiety and computer self-efficacy of students in elementary and secondary schools in Turkey and found that the overall mean of computer anxiety scores was moderate (M=87.89). Contrary to the finding of this study, Singh et al. (2019) investigated the computer anxiety of veterinary science students of Punjab, India and found that the majority of the students possessed low level of computer anxiety.

The finding of this study also revealed that there is no significant difference in the computer anxiety of male and female junior secondary school students in Nigeria. The mean score of male respondents is 22.34 while the mean score of the female respondents is 21.88, the mean computer anxiety score of male students is slightly higher than that of their counterparts. The higher deviation around the mean of the male students revealed that the performances of the male students are not as uniform as the female students that is, the entire female students have similar performances as opposed to the male students. This explains the reason why the male students' better performances are not significant because the sets of male students with good performances and the sets with bad performances did so most likely due to certain variables which are treatments the students are exposed to, which necessitated the reason this study measures the gender performances. The result of this study is in agreement with the findings of Olatoye (2009), who investigated the influence of computer anxiety on computer utilization among senior secondary school students in Ogun state, Nigeria and found that there is no significant difference between male and female students' computer anxiety.

The finding of this study also revealed that there is a significant negative relationship (r = -0.432; p-value = 0.001) between computer anxiety and Computer Studies performance of junior secondary school students' in Nigeria. This means that students who scored higher in Computer Studies examination had lower level of computer anxiety. The finding of this study is in agreement with the findings of Singh et al. (2019), who investigated the computer anxiety of veterinary science students of Punjab, India and found that computer anxiety was negatively correlated with student's grade point.

**CONCLUSION AND RECOMMENDATIONS**

Results indicated that most of the students used in the study were mildly anxious when dealing with computer; and there is a significant negative relationship between computer anxiety and Computer Studies performance of junior secondary school students' in Nigeria. Glaister (2007) found that people who reported to have medium to high levels of computer anxiety performed less well than those with low level of computer anxiety in an examination involving the use of computers. Therefore, students can improve by constant practice with computers. When the students did not upgrade their knowledge of computer usage, it will lead to fear and afraid of using the computer application. When this situation happened, they easily think that every step will lead to mistakes in using the computer because they do not have sufficient knowledge about the latest computer application. The study also revealed that there is no significant difference in the computer anxiety of male and female junior secondary school students in Nigeria. This implies that there are no longer distinguishing cognitive, affective and psychomotor skill achievements of students in respect of gender.

The recommendations and suggestions of this study were made so that the school principals (managers), Computer Studies teachers can take advantages from these findings in order to make their students more capable of handling computer and reduces anxious among their students. First, the school principals should provide effective computer training to each of their students to make sure that they are not anxious in handling computer. An effective training can help to increase the students' knowledge, skills and ability. Second, government should provide enough computer sets to schools so that each students will be entitled to a computer during their training. According to Olatoye (2011), level of computer training however has a positive relationship with computer utilization. Third, government, Ministry of Education and school principals should endeavor to organize seminars, conferences, training and re-training for Computer Studies teachers so that they can update and upgrade their knowledge and improve in their Computer Science teaching methods. Fourth, Computer Studies' teachers should indulge the students in constantly using computer for assignments and web-browsing so as to enhance their confidence and reduce anxiety in Computer Studies classroom. Fifth, students should frequently use computers to avoid anxiety. Further study is needed to replicate the current study with Senior Secondary School students for comparative studies.

**Author's Biography**

Dr. (Mrs.) Chinelo Blessing Oribhabor is a lecturer at the University of Africa, Toru-Orua, Bayelsa State, Nigeria. She specializes in Test, Educational Measurement and Evaluation, Research Methods.